# Structural Changes Related to the Magnetic Transitions in Hexagonal $InMnO_3$

T. Yu[1], T. A. Tyson[1,*], P. Gao[1], T. Wu[1], X. Hong[2] and S. Ghose[2], and Y.-S. Chen[3]

[1]Department of Physics, New Jersey Institute of Technology, Newark, NJ 07102
[2]National Synchrotron Light Source, Brookhaven National Laboratory, Upton, NY 11973
[3]ChemMatCARS , University of Chicago and Advanced Photon Source, Argonne National Laboratory, IL 60439.

*Corresponding Author: T. A Tyson, e-mail:tyson@njit.edu

## Abstract

Two magnetic ordering transitions are found in $InMnO_3$, the paramagnetic to antiferromagnetic transition near ~118 K and a lower possible spin rotation transition near ~42 K. Multiple length scale structural measurements reveal enhanced local distortion found to be connected with tilting of the $MnO_5$ polyhedra as temperature is reduced. Strong coupling is observed between the lattice and the spin manifested as changes in the structure near both of the magnetic ordering temperatures (at ~42 K and ~ 118 K). External parameters such as pressure are expected to modify the coupling.

# I. Introduction

To understand the coupling of the lattice with the spin degrees of freedom in $InMnO_3$ and the general hexagonal $RMnO_3$ systems, detailed temperature dependent pair distribution function (PDF), single crystal diffraction, and XAFS measurements were conducted. These measurements reveal strong coupling manifested as changes in the lattice parameters near $T_N$ (~120 K) and near a possible spin rotation transition, $T_{SR}$ (~40 K). The PDF and single crystal measurements reveal enhanced tilting of the $MnO_5$ polyhedra as temperature is reduced. The results suggest that tuning the crystal structure with pressure or strain can modify the magnetic transition temperature and possibly its coupling to ferroelectricity in these materials. The study provides details on the coupling between spin and lattice in the broader class of $RMnO_3$ systems.

In this specific class of materials the transition to the ordered ferroelectric state ($T_{FE}$) occurs between ~800 and ~1200 K while the ordered magnetic states occur at significantly lower temperature ($T_N$~75) [1]. This hexagonal structure can also be stabilized in large radius cation systems by quenching them from high temperature or by depositions on substrates which induce strain. Evidence of structural changes at the magnetic ordering transition temperatures has been seen in both bulk and single crystal structural measurements.

Anomalies in the dielectric constants, the linear expansion coefficients and phonon frequencies suggest a coupling between the magnetic and ferroelectric order at low temperature [2,3] in $HoMnO_3$. Sharp features are observed at $T_{SR}$ (spin rotation temperature, corresponding to in-plane rotation of Mn spins near ~40 K) and $T_{Ho}$ (Ho moment ordering near 10 K) in addition to the paramagnetic to antiferromagnetic ordering transition near ~80 K. The local structure of $HoMnO_3$ was studied in detail by X-ray absorption spectroscopy [4]. Local structural measurements on hexagonal $HoMnO_3$ show that the transition from the paramagnetic to the antiferromagetic phase near ~70 K is dominated by changes in the a-b plane Mn-Mn bond distances. It is argued that the spin rotation transition near ~40 K involves both Mn-Mn and nearest neighbor Ho-Mn interactions while the low temperature transition below 10 K involves all interactions, Mn-Mn, Ho-Mn (nearest and next nearest) and Ho-Ho correlations. Complementary DFT calculations in that work reveal asymmetric polarization of the charge density of Ho, O3 and O4 sites along the c-axis in the ferroelectric phase. This polarization facilitates coupling between Ho atoms on neighboring planes normal to the c-axis. Neutron pair distribution function measurements on $LuMnO_3$ [5] reveal a reduction in space group symmetry from $P6_3cm$ to $P6_3$ concomitant with the appearance of local distortions. The distortions are characterized by splitting in the Mn-O-Mn angles with enhanced separation between distinct in-plane Mn-O-Mn bond and enhanced polyhedral tilting angles with lowering of temperature.

It is generally argued that the transition near ~40 K is due to the coupling of the Mn 3d and R site 4f magnetic moments. However, hexagonal systems with no 4f electrons, such as nanoscale $LuMnO_3$ [6], exhibit as spin transition near 40 K. To understand the true nature of the coupling and structural changes with temperature in $InMnO_3$ (and the general $RMnO_3$ system), detailed single crystal diffraction measurements for very high resolution atomic position determination, PDF measurements for local and intermediate range structural measurements and XAFS measurements for local site specific structural studies have been conducted. Heat capacity and magnetic susceptibility measurements are used to identify the magnetic transitions

## II. Experimental Methods

Single crystals of hexagonal $InMnO_3$ were prepared as given in our previous study [7]. Diffraction measurements on the $InMnO_3$ crystals were conducted at the Advanced Photon Source (Argonne National Laboratory) beamline 15-ID-B with a wavelength of 0.41328 Å. Refinement of the data was done using the program Olex2 [8] after the reflections were corrected for absorption (see Ref. 7). For pair distribution function (PDF) measurements powder samples were ground from the single crystals to 500 mesh size. Experiments were conducted at beamline X17B3 at Brookhaven National Laboratory's

National Synchrotron Light Source (NSLS). The wavelength was set at 0.152995 Å and data were measured using a Perkin Elmer detector with the sample to detector distance of 255.33 mm. $Q_{max}$ = 26 $Å^{-1}$ was used in data reduction. The methods utilized for analysis of the PDF data are described in detail in Ref. [9]. For the fits in R-space, several ranges were chosen: $1.2 < R < r_{max}$ ($r_{max}$= 15 Å (short range structure) and 60 Å (intermediate range structure)). For XAFS measurements, polycrystalline samples were also prepared from the single crystals by grinding and sieving the material (500 mesh) and brushing it onto Kapton tape. Layers of tape were stacked to produce a uniform sample for transmission measurements with jump $\mu t \sim 1$. Spectra were measured at the NSLS beamline X3A. Measurements were made on warming from 30 K to 300 K in a sample attached to the cold finger of a cryostat. Three to four scans were taken at each temperature. The uncertainty in temperature is < 0.2 K. At the Mn K-Edge, a Mn foil reference was employed for energy calibration. The reduction of the X-ray absorption fine-structure (XAFS) data was performed using standard procedures [10]. For the Mn K-Edge data, the k-range 1.56< k < 12.53 $Å^{-1}$ (k=($\sqrt{(E-E_0)2m}/\hbar$) and the ionization energy is $E_0$) and the R-range 0.71 < R < 3.64 Å were used with $S_0^2$ = 0.90. Coordination numbers for the atomic shells were fixed to the crystallographic values. The limited energy range at the Mn K-edge constrained the modeling to the shells: <Mn-O>, Mn-In(short), Mn-Mn and Mn-In(long). The temperature dependence of the bond Debye-Waller factors ($\sigma^2$) was modeled by static contribution ($\sigma_0^2$) plus a single parameter ($\theta_E$) Einstein model using the functional form $\sigma^2(T)=\sigma_0^2+\frac{\hbar}{2\mu k_B\theta_E}\coth(\frac{\theta_E}{2T})$ [11], where $\mu$ is the reduced mass for the bond pair. This simple model represents the bond vibrations as harmonic oscillations of a single effective frequency proportional to $\theta_E$. It provides an approach to characterize the relative stiffness of the bonds. For heat capacity and magnetic susceptibility measurements, a Quantum Design Physical Properties Measurements system was utilized.

# III. Results and Discussion

### III. a. Heat Capacity and Magnetic Measurements

Figure 1 shows the crystal structure of $InMnO_3$ indicating the Mn sites at the center of the $MnO_5$ polyhedra with out of plane O (O1 and O2) and in-plane O (O3 and O4) oxygen atoms labeled in addition to the In and Mn sites (See Ref. 7). The structure is similar to the small ion hexagonal $RMnO_3$ systems with buckling of the In planes at low temperature (below the ferroelectric ordering temperature). This buckling also coincides with tilting of the $MnO_6$ polyhedra, defined by the O1-O2 vector for each polyhedron, relative to the c-axis. The bonding in the $MnO_6$ polyhedra is highly ordered compared to the systems with other closed shell systems R-site systems ( R=Y, Sc and Lu [12]). We find for $InMnO_3$ <Mn-O> = 1.949(2) Å, compared to $ScMnO_3$ (<Mn-O> = 1.932(3) Å), $YMnO_3$ (<Mn-O> = 1.984(14) Å), and $LuMnO_3$ (<Mn-O> = 1.966(8) Å), the coordination of Mn by O revealed by the bond distance is closest to $ScMnO_3$. In addition the distribution of Mn-O bonds is significantly more narrow (more ordered) for $InMnO_3$ than any of these other systems and the neighbor Mn-Mn distribution is the smallest for $InMnO_3$ compared to the R= Y, Lu and Sc systems.

Heat capacity measurements between 2 and 300 K were conducted and are as shown in Fig. 2(a). There are two clear transitions, (1) the paramagnetic to antiferromagnetic transition, $T_N$, near 118 K and (2) a weaker but clearly visible feature near ~42 K (inset). Susceptibility measurements are given in Fig. 2(b) at 2 T and at 0.1 T (inset). In the susceptibility measurements at low field there is a discernable change near ~42 K showing that the second transition is clearly of magnetic origin. This feature near ~43 K typically seen in system with 4f electrons at the R-site ($RMnO_3$) such as $HoMnO_3$ indicating that the spin rotation $T_{SR}$ can be driven by Mn magnetic sites only depending on the details of the crystal structure such and Mn-Mn bond distance. The same feature was also see in $LuMnO_3$ prepared as a nanoscale material but not in bulk form [6]. Extrapolation of the inverse susceptibility using the Curie-Weiss law for the paramagnetic phase (using data between 175 and 320 K) for the 2 T data yielded a value of the Curie-Weiss temperature of $\theta_p$ = -379 K (Figure 2(b)). )). This value should be compared to values of -417 K, -519 K and -495 K found for hexagonal $YMnO_3$ , $LuMnO_3$ and $ScMnO_3$, respectively [13]. Not that the ratio f = $-\theta_p/Tc$ defines a so-called frustration parameter [14]. The value of f ~ 3.2 (379/118) for hexagonal $InMnO_3$ is lower than that found for $YMnO_3$ (~5.9), $LuMnO_3$ (~5.8) and $ScMnO_3$ (~3.8) showing that this compact system has reduced frustration and a stable magnetic ordering of the Mn magnetic layer. Detailed temperature dependent structural studies were conducted to understand the underlying short range to long range atomic order accompanying the stable magnetic states.

### III. b. Structural Measurements

The room temperature $P6_3cm$ structure was used as a model for all temperatures and refined against a goodness of fit parameter $R_W$ ($R_W = \left\{\frac{\sum_{i=1}^{N} w(r_i)[G_{Obs}(r_i)-G_{Calc}(r_i)]^2}{\sum_{i=1}^{N} w(r_i)[G_{Obs}(r_i)]^2}\right\}$ ). Note that G(r) is the reduced pair distribution function which oscillates about zero and is obtained directly from the scattering data, S(Q). $G(r) = \frac{2}{\pi}\int_0^\infty Q[S(Q)-1]\sin(Qr)dQ$ is related directly to the standard pair distribution function. The goodness of fit parameter varies continuously with temperature (Fig. 3) and was compared over the whole temperature range for to fitting regions: 1.2 Å $<r_{max}<$15 Å (short range structure) and 1.2 Å $<r_{max}<$60Å (intermediate range structure). Examining the variation of Rw with temperature (80 to 300 K) reveals that the fits improve for the intermediate range structure fits as temperature is reduced and the anomalous points near the Neél temperature, $T_N$~118 K. The variation in Rw over this temperature range is ~11 %. On the other hand the short range fits get worse with temperature for this reduced temperature with variation of ~13%. The results indicate the presence of local distortions which are increasingly (see $r_{max}$=15 Å data) enhanced as temperature is reduced. Typical G(r) reduced radial distribution curves are shown in Fig. 6 focusing on a limited range of r, where the peaks corresponding to the <Mn-O>, <In-O>, <O-O> and <Mn-Mn>/<Mn-In> bonds can be clearly identified. No significant changes are seen in the

<Mn-O> and <In-O> peaks with temperature on crossing the Neel temperature, $T_N$~118 K. However, the <O-O> bonds within the $MnO_6$ polyhedra show perceptible temperature dependence suggesting changes in shape of the $MnO_5$ polyhedral which preserver the bond distances, most likely local distortions. The peak height (Fig. 5) as a function of temperature of the <O-O> peaks shows an increase near $T_N$, indicating that the structural change is linked to the magnetic ordering. This is consistent with the anomalies seen in the fits parameter of the PDF for $r_{max}$ = 60 Å (Fig. 3).

While the temperature dependence of isotropic atomic displacement parameters (ADPs, U parameters) for the heavy atoms are is well behaved (Fig. 6), the curves for the oxygen atoms such as O3 and O4 have anomalous temperature dependence (Fig. 7). It is also found that the atomic positions appear to change continuously with temperature.

The details of the bond distances can also be explored. The Mn-Mn in-plane distances are reduced moving toward a more symmetric triangular lattice as temperature is decreased but still at the lowest temperature (with distinct Mn-Mn bonds). While the average Mn-O distance does not change with temperature the components split or merge. The local structural model ($r_{max}$=15Å, Fig. 9) shows large differences between the Mn-O1 and the Mn-O2 as temperature is reduced without much change occurring with the in plane Mn-O3/Mn-O4 bonds. The In-O bonds are found to show much smoother behavior.

The nature of the local distortions can be further explored by examining carefully R-space fits with $r_{max}$= 60, 15 and 5 Å and looking at the G(r) function near ~4.3 Å (~4.3 Å) as shown in Fig. 10. What is seen is that there is an enhanced splitting of the peak near ~4.3 Å as temperature is reduced (Fig. 10 panel (d)). A low-R shoulder becomes strongly enhanced as temperature is reduced. This peak corresponds to a long Mn-O1 distance. Specifically, it corresponds to the distance between Mn in one layer and O1 in the $MnO_5$ polyhedra in the layer below or above (see Fig. 11). The enhanced splitting is due to increased tilting of the $MnO_5$ polyhedra as temperature is reduced. This is the primary distortion which occurs with reduced temperature. The distortion is short ranged and can only be modeled by fitting for a short region of r-space as seen in Fig. 10. This distortion is also reflected as increased deviation of the data from the high-symmetry model $P6_3cm$ in the short range fits as temperature is reduced (Fig. 3).

To address the structural changes at low temperature in terms of the accurate long range structure, single crystal measurements were conducted. Full refinements of single crystal data were carried out between 10 K and 60 K. Figure 12 shows that there is a significant lattice response near the transition at ~40 K with dips in both the a and c lattice parameters. A drop in Mn-O4-Mn bond angle and jump in the tilt angle with reduce temperature are also seen to occur (Fig. 13). Our analysis found no significant change in the ADPs for the heavy ions (In and Mn), In-O bond distances or the individual Mn-O bonds near the transiton.. Hence like the transition near $T_N$, this transition is related to the changes of $MnO_5$ polyhedra exclusively. Complementing the single crystal results, the in-plane Mn-Mn bond correlations ($\sigma^2$) for hexagonal InMnO3 and LuMnO3 were probed by XAFS measurements (Fig. 14). The in-plane behavior of the Mn-Mn correlation for $LuMnO_3$ and $InMnO_3$ reveal similar stiffness (same Einstein temperature $\theta_E$ ~200 K) and static structural disorder ($\sigma_0^2$). Suggesting that the presence of the feather at ~40K is related with how the 2D $MnO_2$ planes are constructed. Hence, the spin rotation may be primarily structurally driven (not related to Mn and R site spin coupling).

## IV. Summary

Two magnetic ordering temperatures are found in $InMnO_3$, the paramagnetic to antiferromagnetic temperature near ~118 K and a lower possible spin rotation transition near ~40 K. Multiple length scale structural measurements reveal enhanced local distortions connected with tilting of the $MnO_5$ polyhedra as temperature is reduced. Strong coupling is observed between the lattice and the spin manifested as changes in the structure near both of the magnetic ordering temperature. The results suggest that external parameters such as pressure or strain can modify the coupling between magnetic properties and atomic structure.

## V. Acknowledgments

This work is supported by DOE Grant DE-FG02−07ER46402. Synchrotron powder X-ray diffraction and X-ray absorption data acquisition were performed at Brookhaven National Laboratory's National Synchrotron Light Source (NSLS) which is funded by the U.S. Department of Energy. Single crystal X-ray diffraction measurements were performed at the beamline 15-ID-B, Advanced Photon Source, Argonne National Laboratory. ChemMatCARS Sector 15 is principally supported by the National Science Foundation/Department of Energy under Grant NSF/CHE-0822838. Use of the Advanced Photon Source was supported by the U.S. department of Energy, Office of Science, Office of Basic Energy Sciences, under Contract No. DE-AC02-06CH11357. The Physical Properties Measurements System was acquired under NSF MRI Grant DMR-0923032 (ARRA award).

**Fig. 1.** The crystal structure of hexagonal $InMnO_3$. The atomic positions are ($P6_3cm$ space group): In1 in light red at 2a (0, 0, z), In2 in pink at 4b (1/3, 2/3, z), Mn in grey dots at 6c (x, 0, 0), and O1 at 6c (x, 0, z), O2 and O3 at 2a (0, 0, z) and O4 at 4b (1/3, 2/3, z), in blue .

**Fig. 2.** (a) Heat capacity *vs.* temperature of single crystals reveals a Néel temperature of approximately 118 K. Insert in panel (a) shows the shoulder with a peak near 42 K. The error bars are smaller than the symbols. (b) The DC magnetic susceptibility at 2T and at 0.1 T (inset) revealing that the feature near ~42 K is a magnetic transition.

**Fig. 3.** The $R_w$ factor after refinement of PDF experimental data, the pink curve is the result for short range fitting with $r_{max}$ up to 15 Å, the blue one is the result for intermediate range fitting with $r_{max}$ up to 60 Å.

**Fig. 4.** (a) The temperature dependence of the local structure. The bond positions from structural refinement are used to label the peaks. Insert (i) is the expanded range for Mn-O (at ~1.91 Å) and In-O (at ~2.18 Å) peaks. Inset (ii) is the expanded range for O-O (at ~2.58 Å and 2.82 Å) peaks. (b) The temperature dependence of the local structure near Néel temperature. Insert (i) is the expanded range for Mn-O (at~1.91 Å) and In-O (at ˚ ~2.18 Å) peaks. Inset (ii) is the expanded range for O-O (at ~2.58 Å and 2.82 Å) peaks near Néel temperature.

**Fig. 5.** The PDF peaks height relative to the vale at 300K as a function of temperature for the Mn-O (at~1.91 Å), In-O (at ~2.18 Å), and O-O (at ˚~2.58 Å and 2.82 Å) peaks, showing that there are changes in the widths for O-O bond distribution indicating the distortion in the O-O bonds near the Néel temperature.

**Fig. 6.** The temperature dependent atomic displacement parameters (ADPs) for In1, In2, and Mn, in short range ($r_{max}$ = 15Å) and intermediate range ($r_{max}$ = 60Å). There is no obvious anomaly in the behavior of the ADPs of heavy atoms, such as In1, In2, and Mn.

**Fig. 7.** The temperature dependent atomic displacement parameters (ADPs) for O1, O2, O3, and O4 are compared both in short range ($r_{max}$ = 15 Å) and intermediate range ($r_{max}$ = 60 Å).

**Fig. 8** (The temperature dependent bond distances for Mn-Mn both from the short range and intermediate range.

**Fig. 9.** The temperature dependent bond distances for Mn-O both from the short range and intermediate range.

**Fig. 10.** **(**a) A comparison of the $P6_3cm$ model and the observed PDF data at 300 K with a short range (r = 15 Å) fitting. (b) A comparison of the $P6_3cm$ model and the observed PDF data at 300 K with an intermediate range (r = 60 Å) fitting. (c) and (d) The temperature dependence of the local structure at the peak around 4.2 Å and 4.3 Å.

**Fig. 11.** (a) Bonds and title angles in the $MnO_5$ polyhedral. (b) The Mn-O1 bond distances to the nearest polyhedral in the layer below.

**Fig. 12.** Single crystal results. (a) The lattice parameters, a and c are compared over the temperature range 10-60K, showing that there is anomalous behavior near 40 K. (b) The temperature dependent c/a ratio and volume are compared.

**Fig. 13.** The temperature dependent in-plane angles, Mn-O3-Mn and Mn-O4-Mn and the tilting angle $\alpha$ both from single crystal data.

**Fig. 14.** XAFS derived Mn-Mn bond width parameter $\sigma^2$ *vs.* temperature for hexagonal $InMnO_3$ (a) compared to hexagonal $LuMnO_3$ (b). Note the similarity in the values of the Einstein ($\theta_E$) temperature and static disorder parameters ($\sigma_0^2$).

Fig. 1.   Yu *et al.*

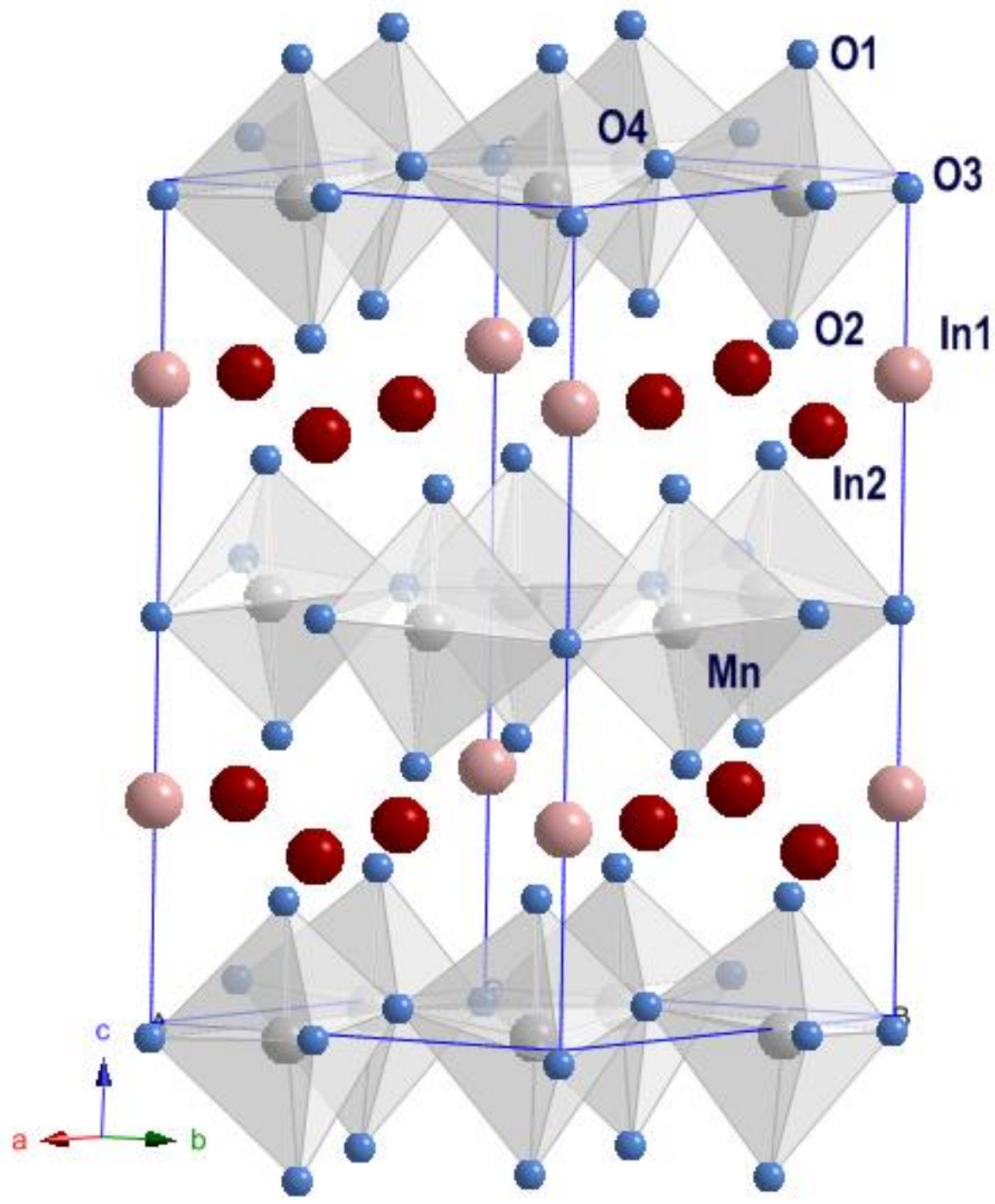

+

Fig. 2. Yu *et al.*

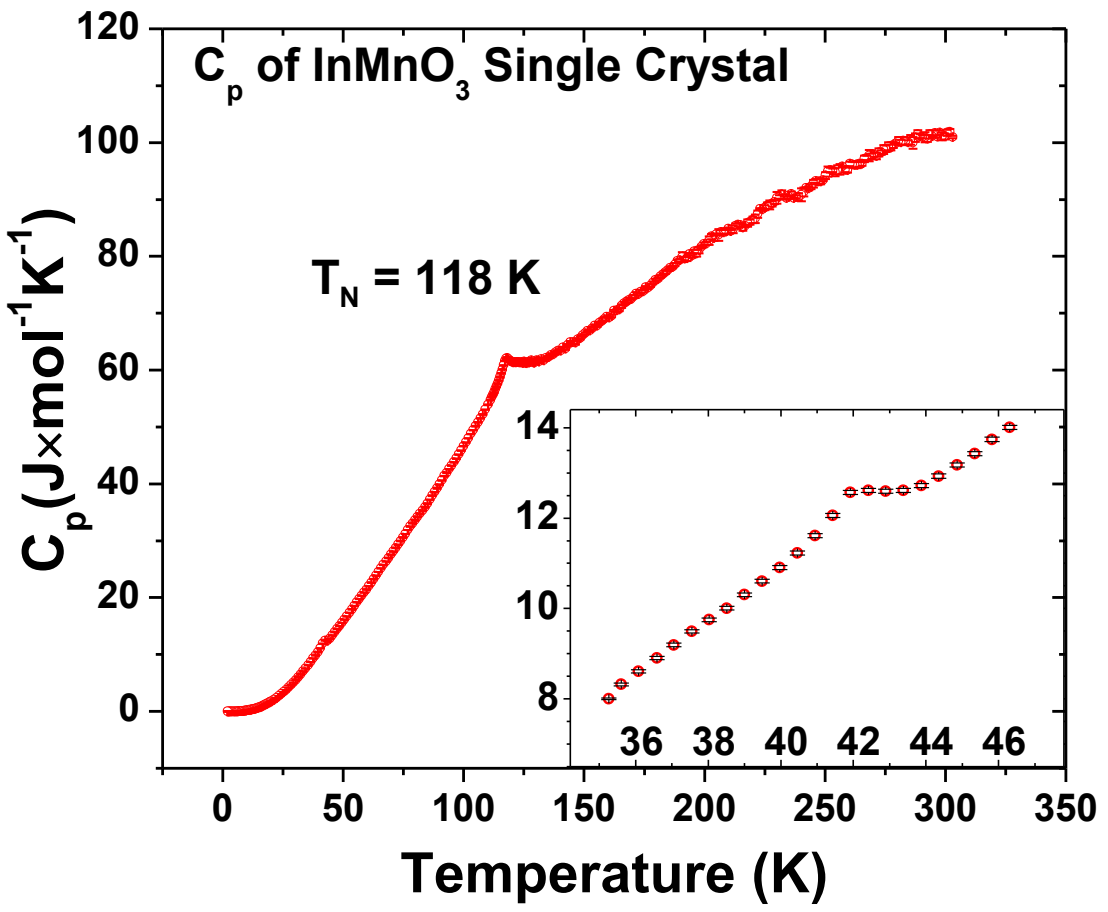

(a)

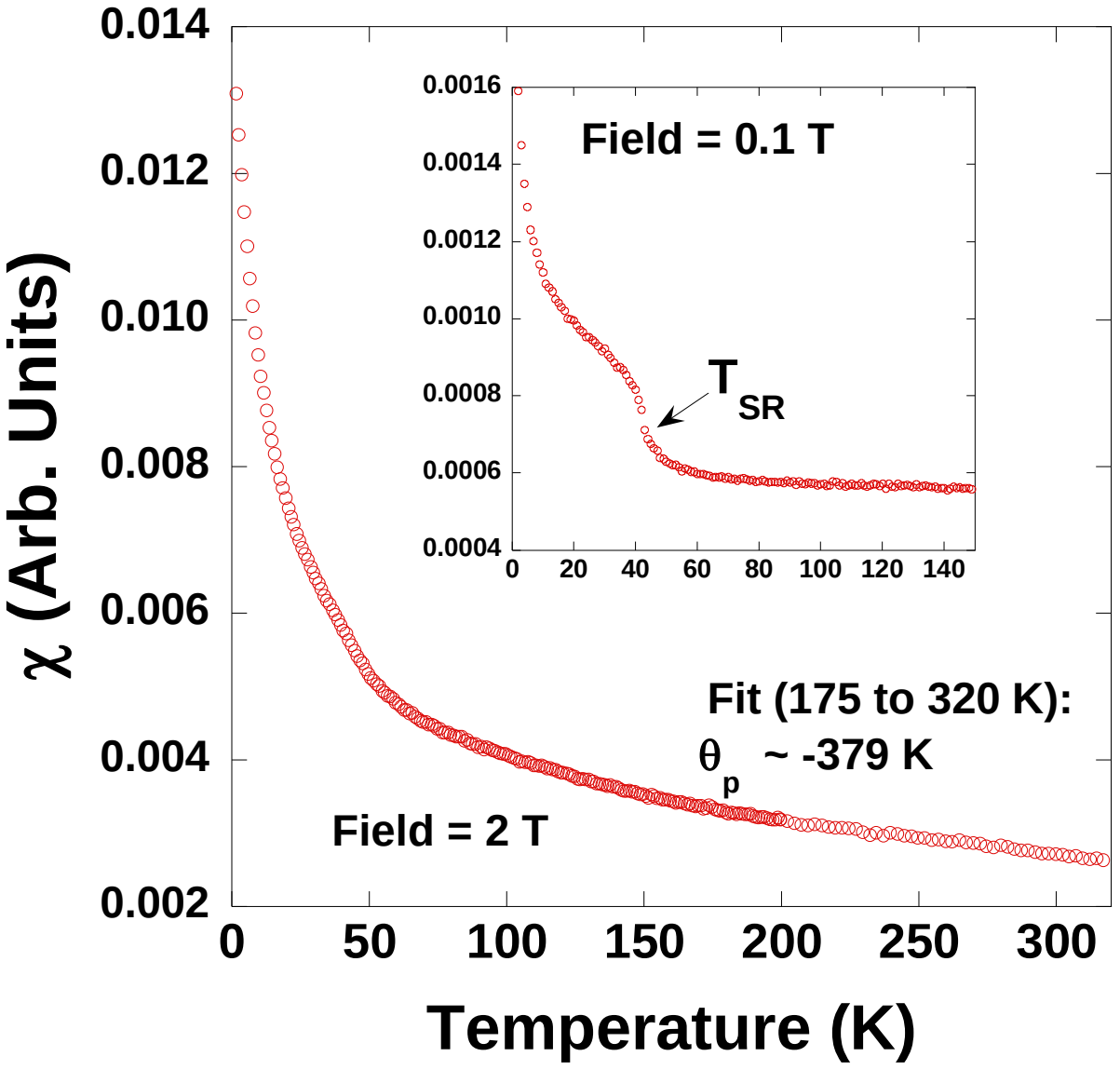

(b)

Fig. 3.   Yu *et al.*

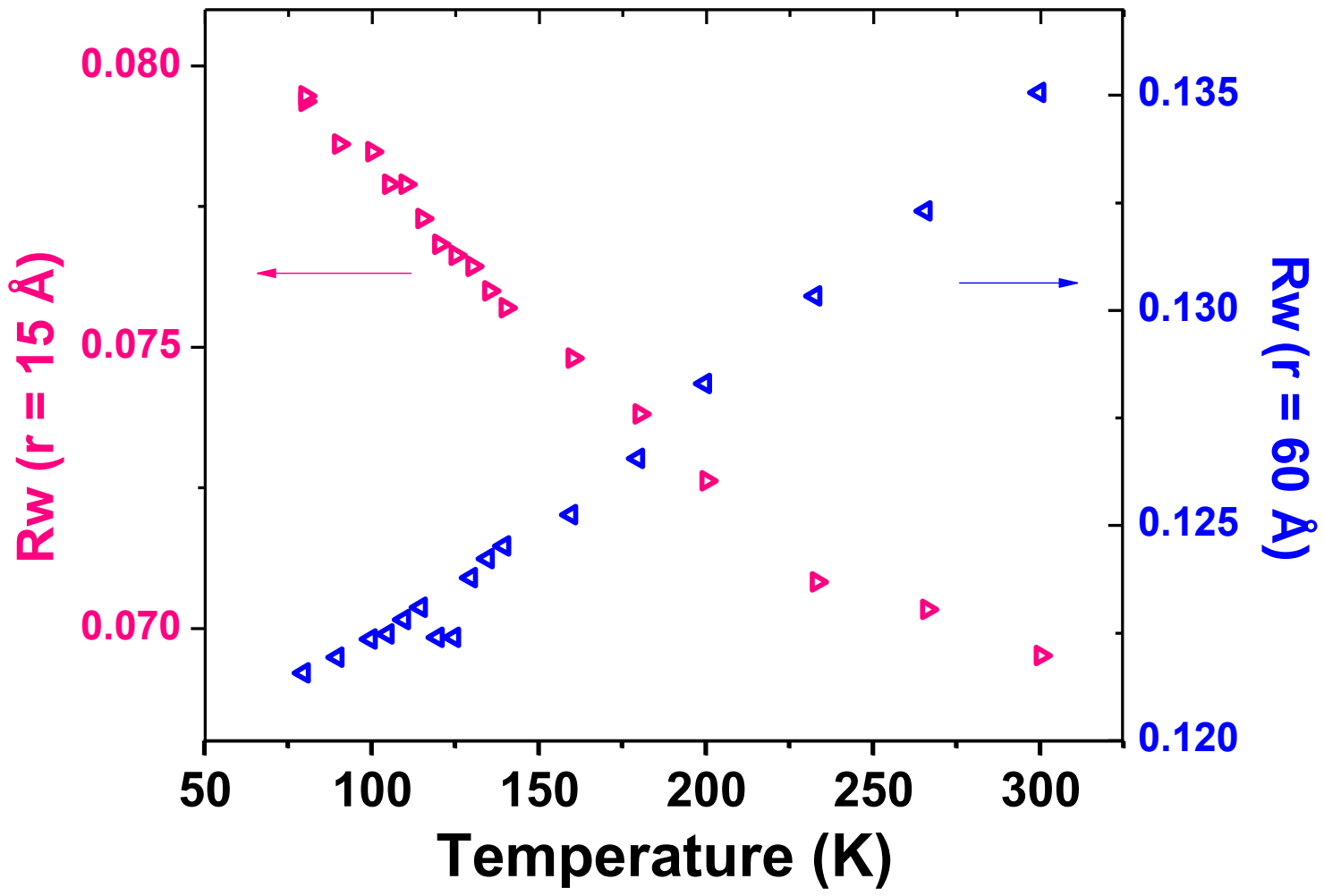

Fig. 4. Yu *et al.*

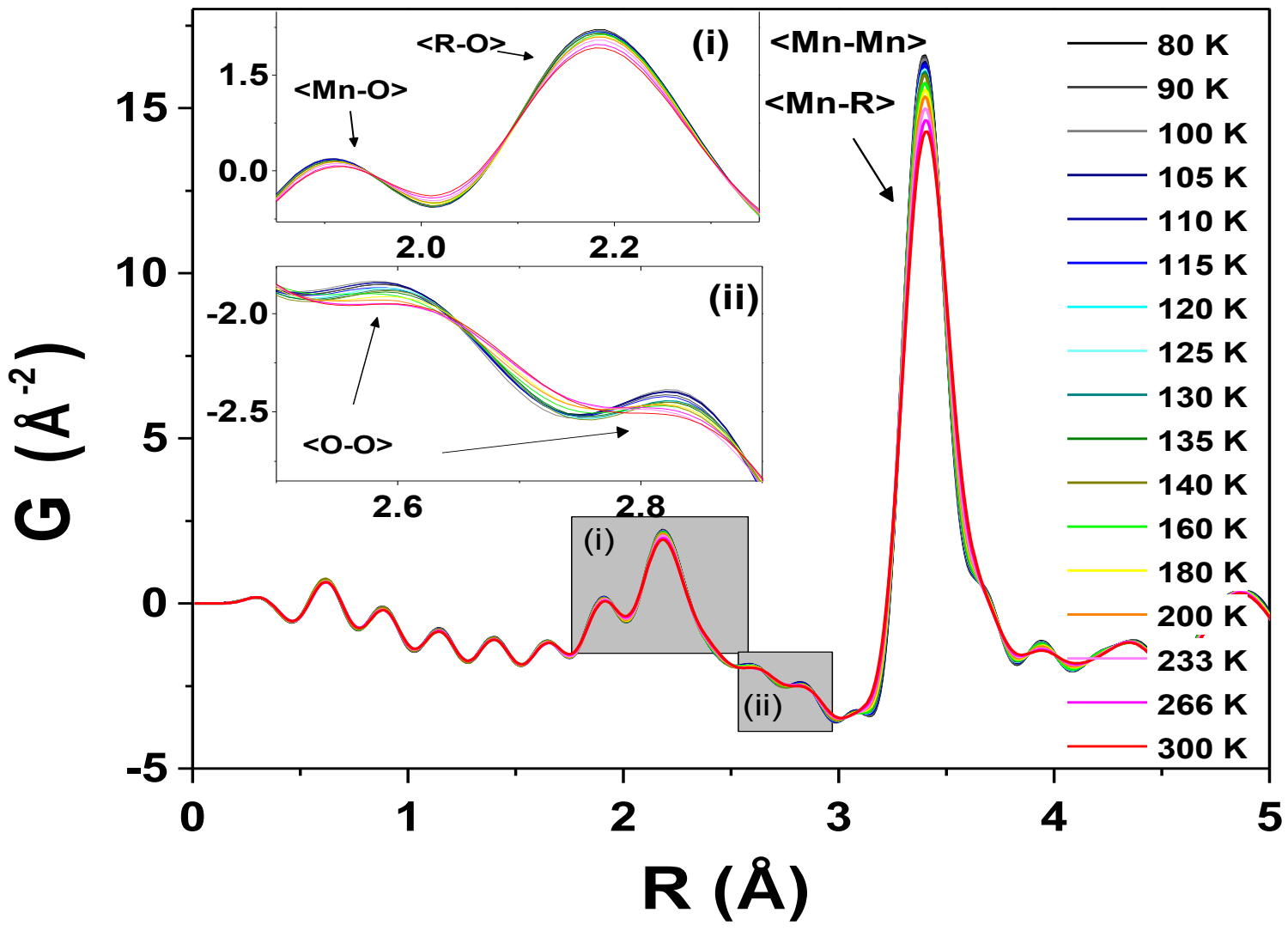

(a)

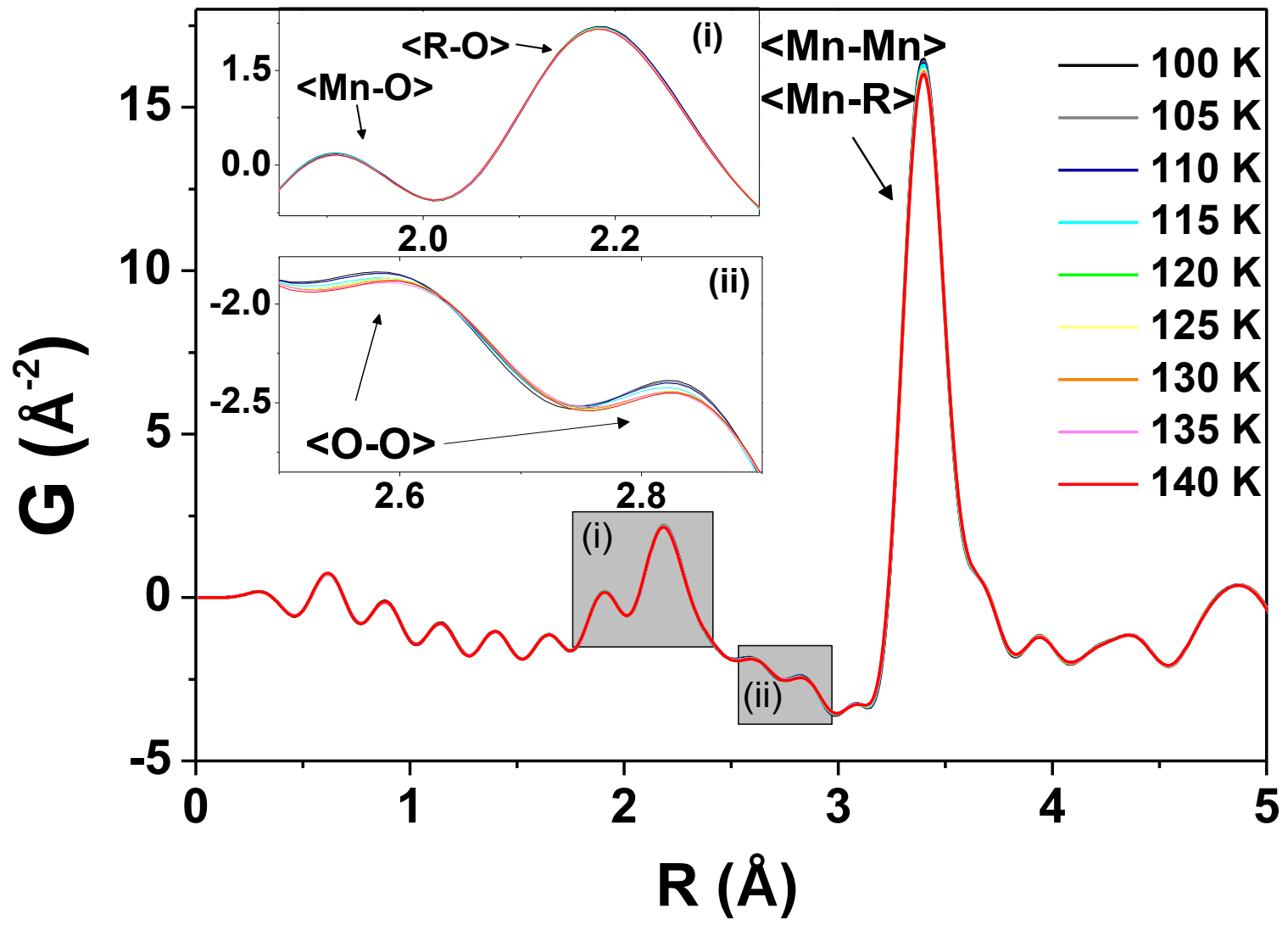

(b)

Fig. 5. Yu *et al.*

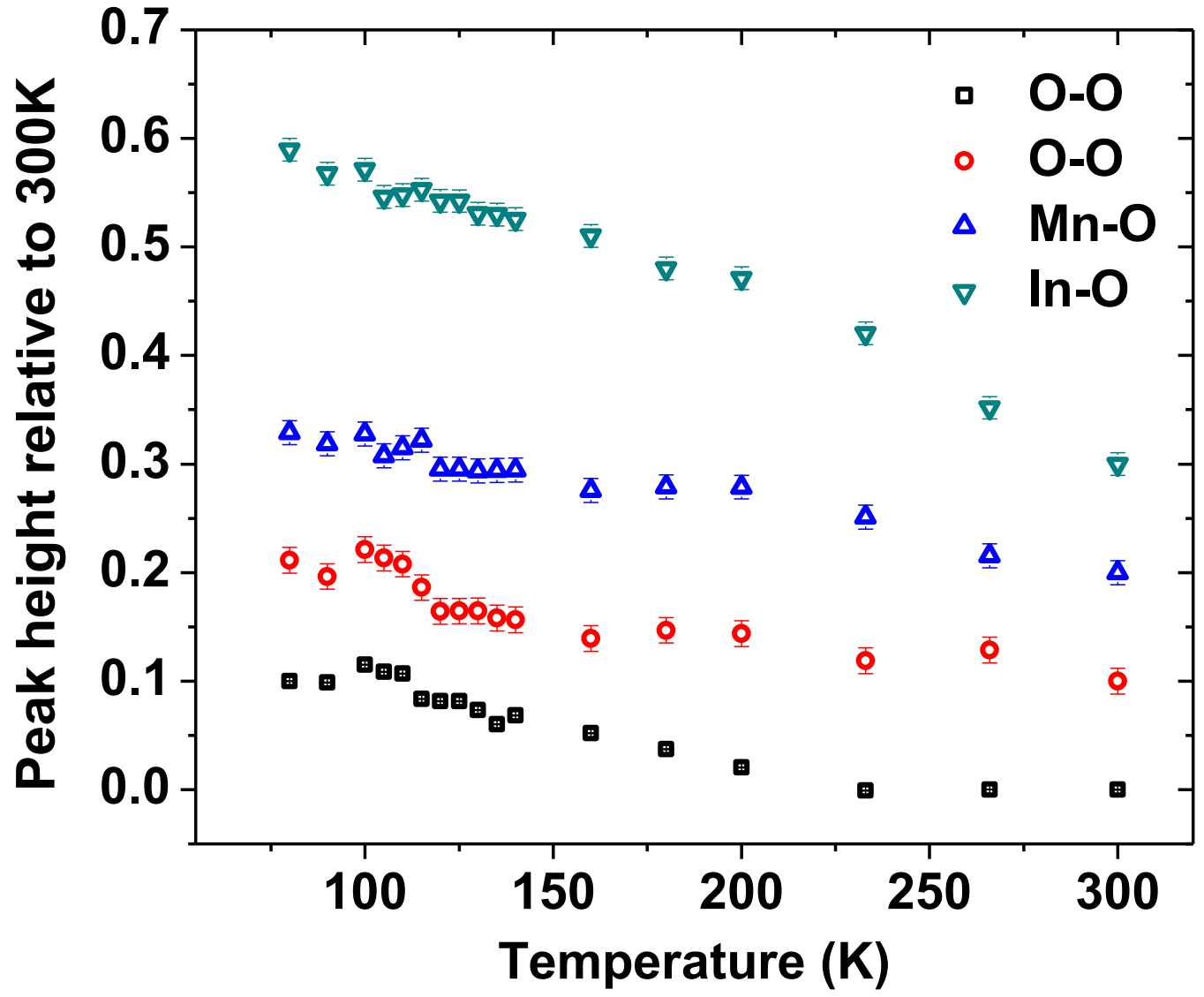

Fig. 6. Yu *et al.*

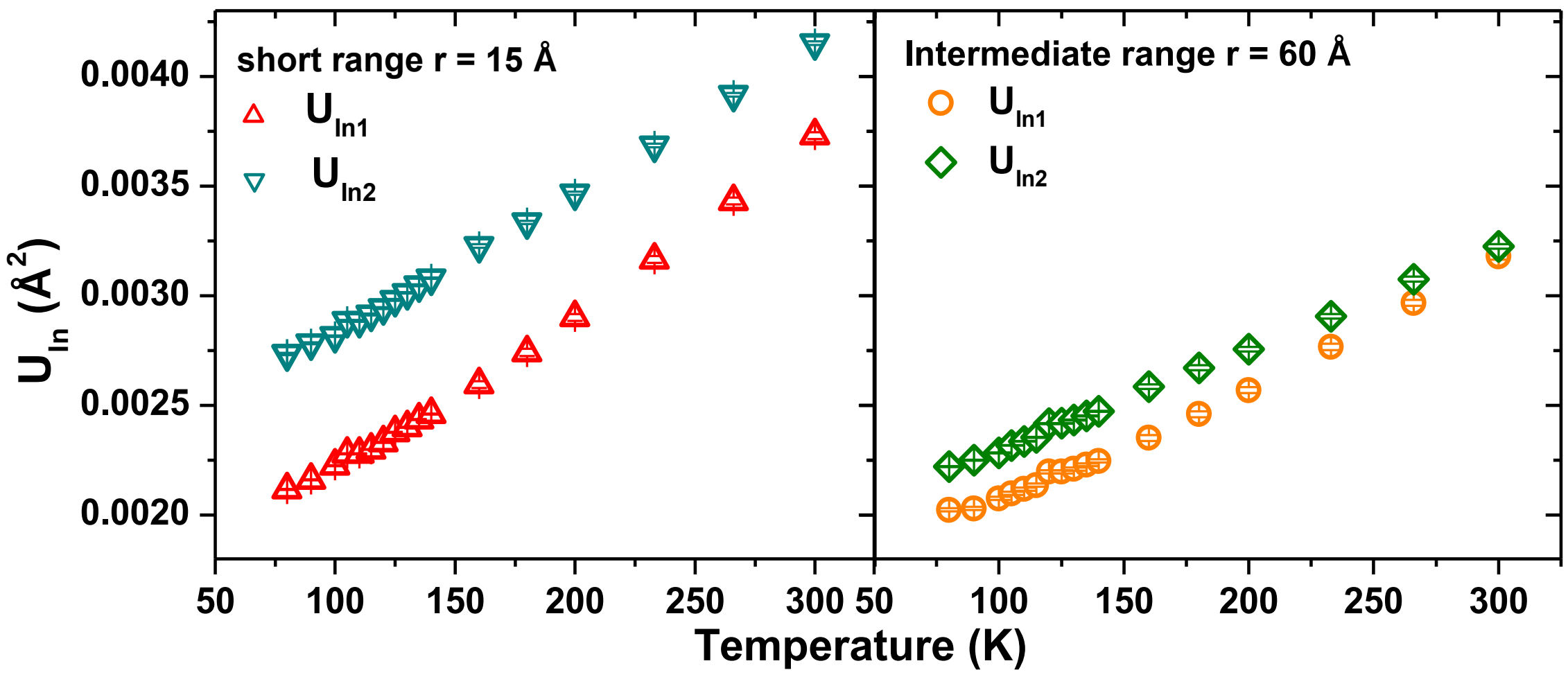

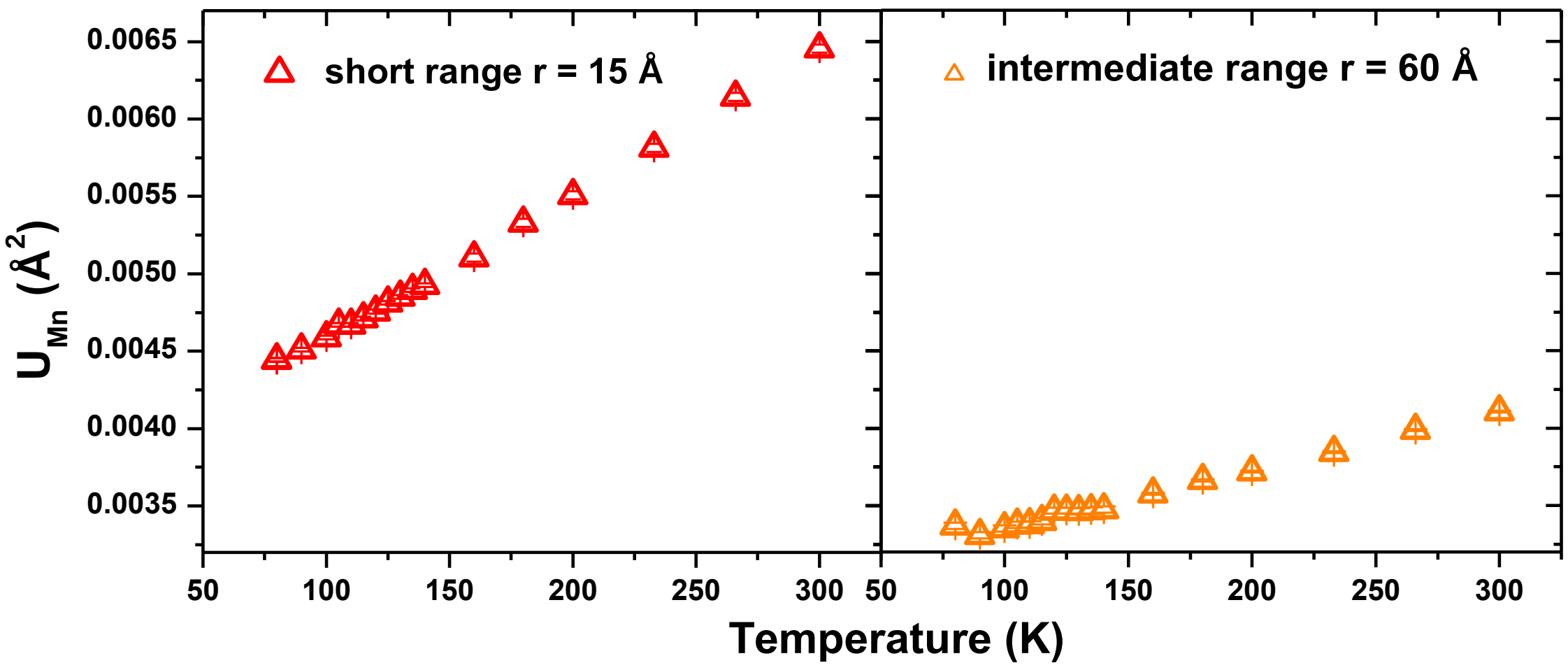

Fig. 7. Yu *et al.*

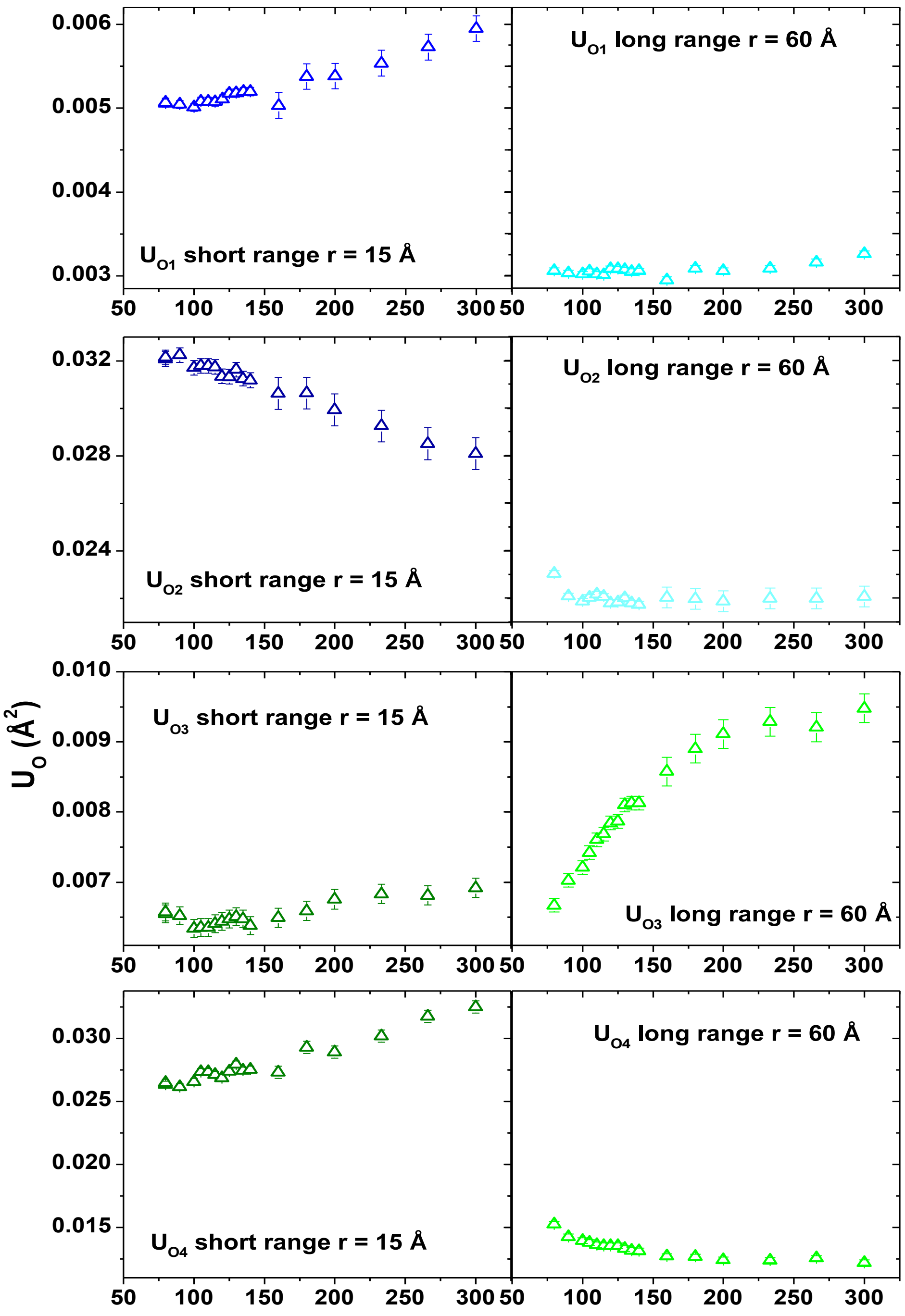

U$_{O1}$ short range r = 15 Å
U$_{O1}$ long range r = 60 Å
U$_{O2}$ short range r = 15 Å
U$_{O2}$ long range r = 60 Å
U$_{O3}$ short range r = 15 Å
U$_{O3}$ long range r = 60 Å
U$_{O4}$ short range r = 15 Å
U$_{O4}$ long range r = 60 Å
U$_{O}$ (Å$^2$)
0.006
0.005
0.004
0.003
0.032
0.028
0.024
0.010
0.009
0.008
0.007
0.030
0.025
0.020
0.015
50
100
150
200
250
300

Fig. 8. Yu *et al.*

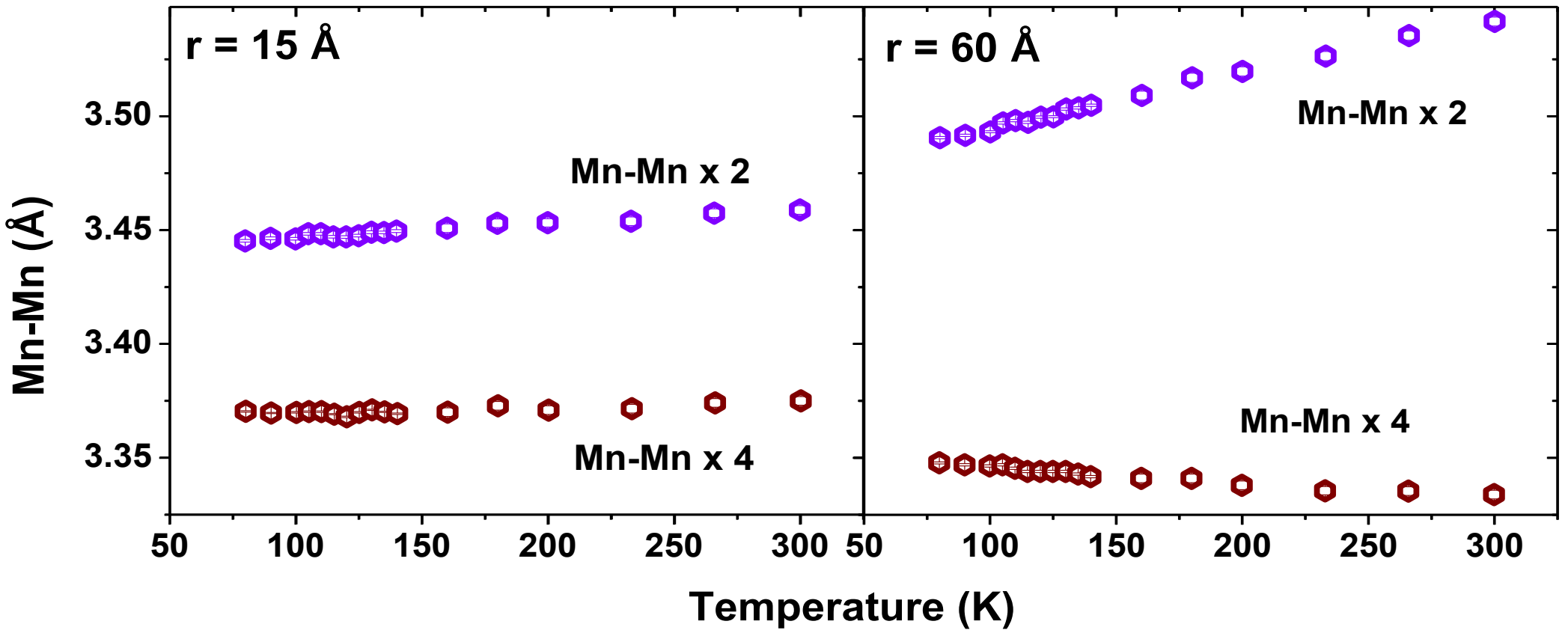

Fig. 9. Yu *et al.*

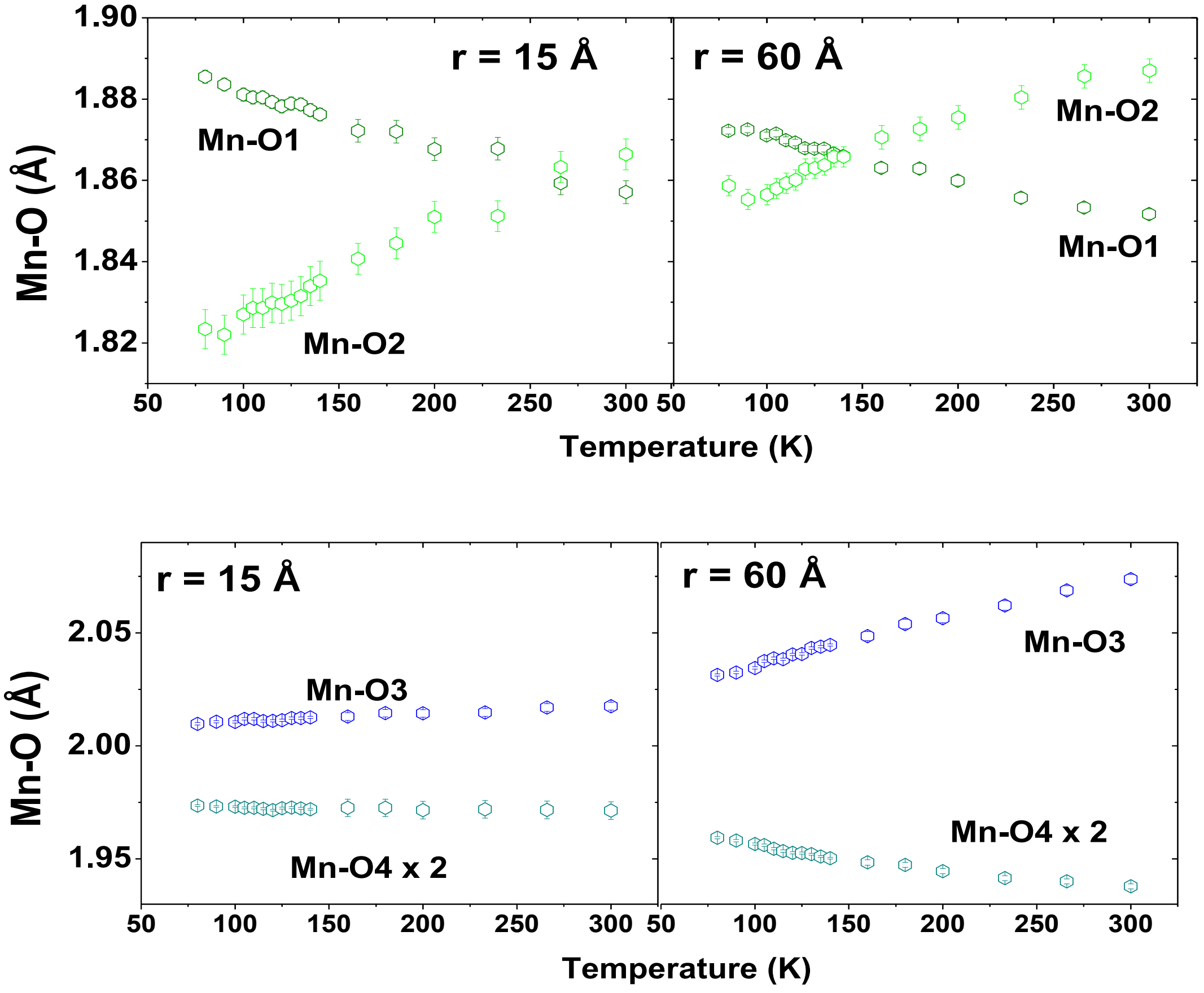

r = 15 Å
r = 60 Å
Mn-O1
Mn-O2
Mn-O2
Mn-O1
Mn-O (Å)
Temperature (K)
r = 15 Å
r = 60 Å
Mn-O3
Mn-O4 x 2
Mn-O3
Mn-O4 x 2
Mn-O (Å)
Temperature (K)

Fig. 10. Yu *et al.*

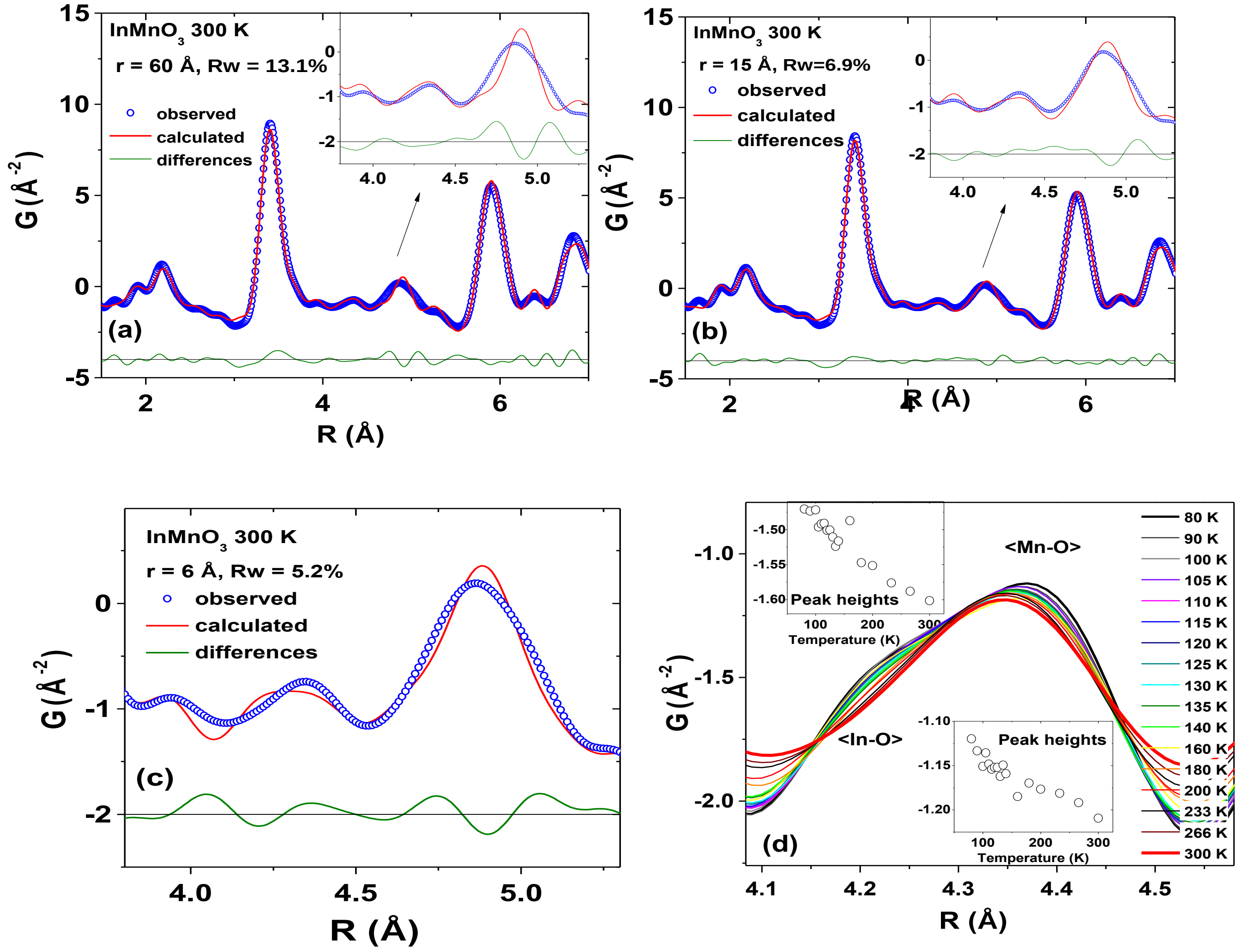

Fig. 11. Yu *et al.*

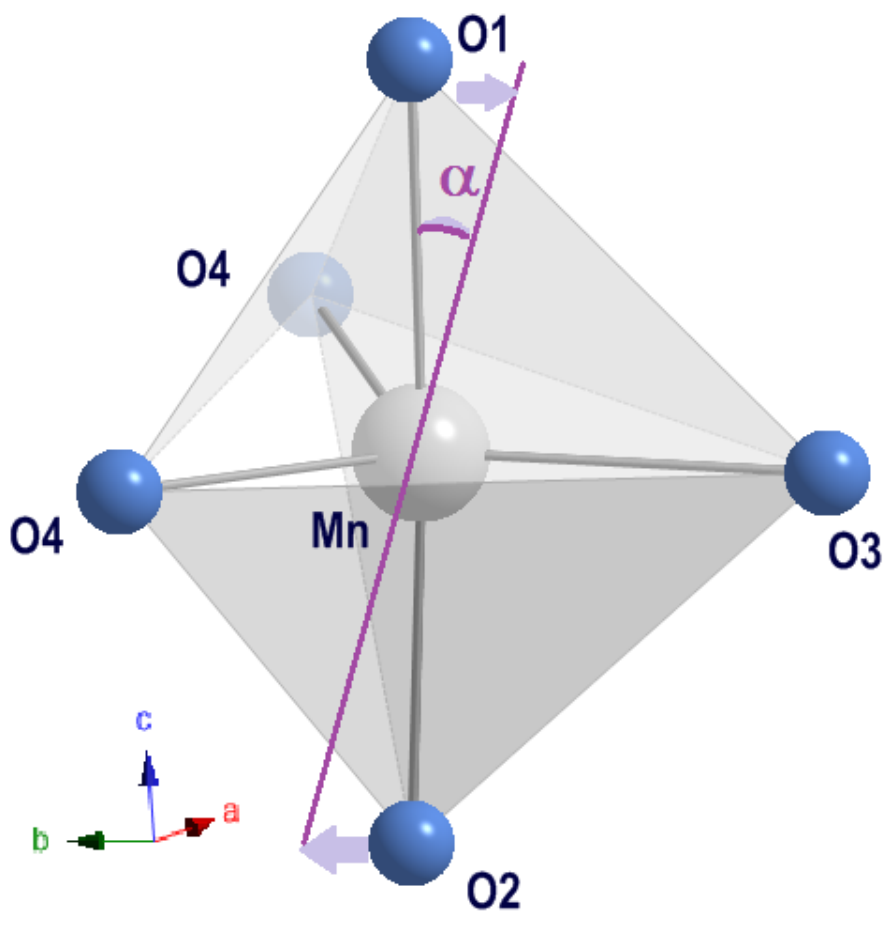

(a)

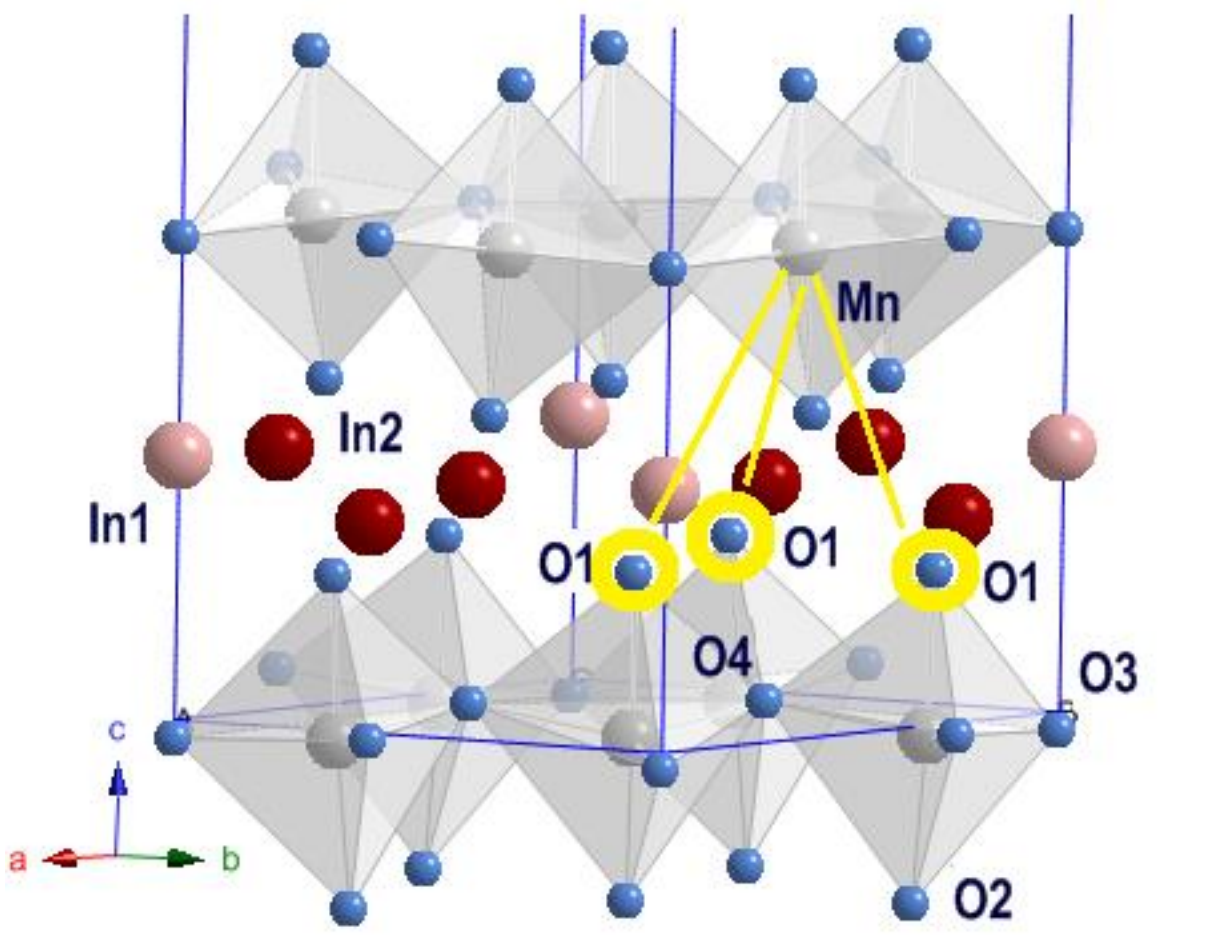

(b)

Fig. 12. Yu *et al.*

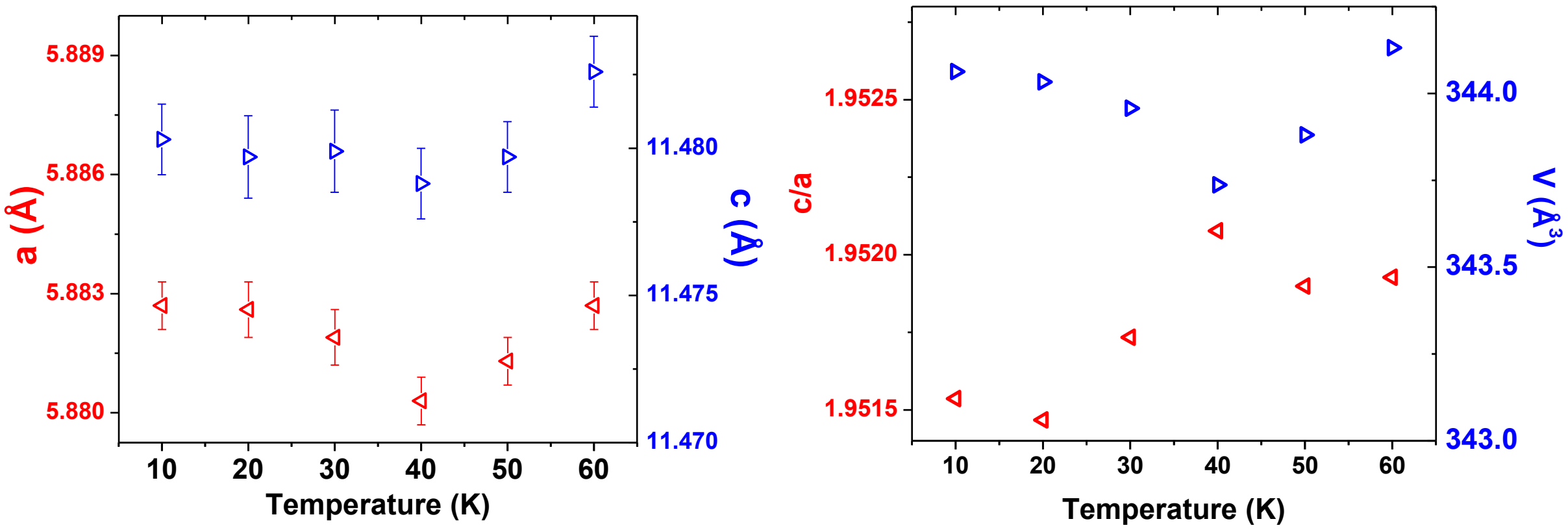

Fig. 13.   Yu *et al.*

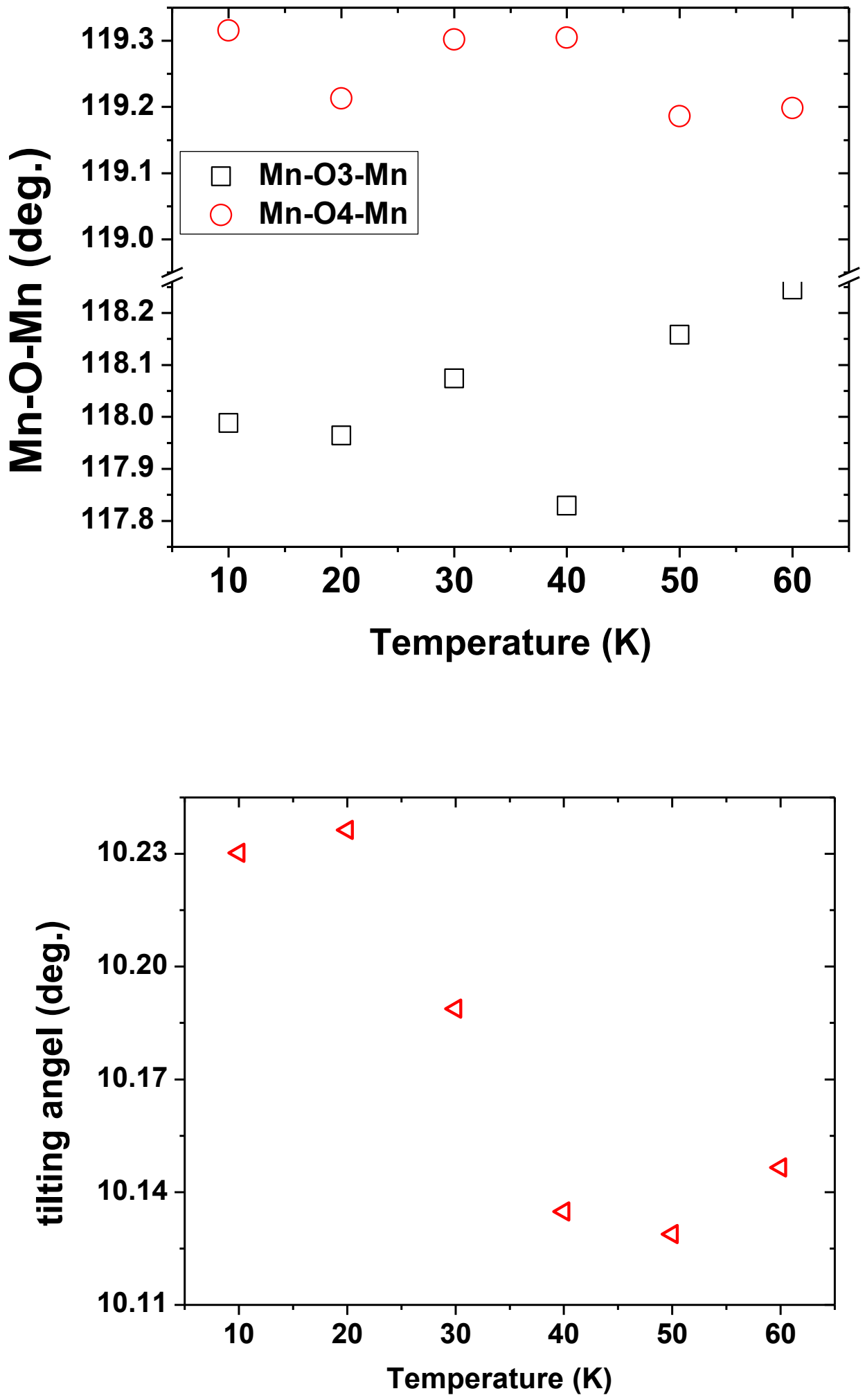

Fig. 14. Yu *et al.*

(a)

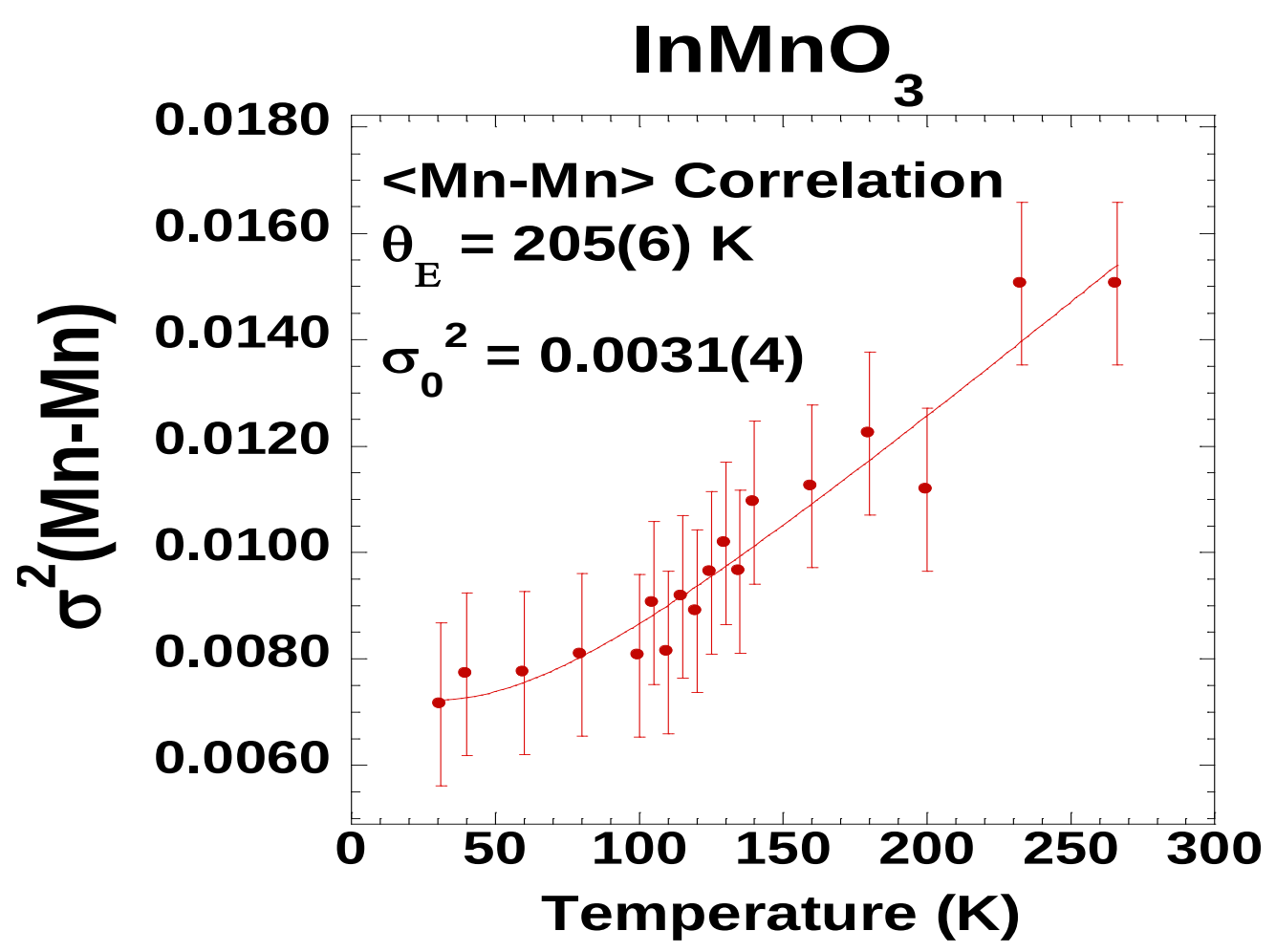

(b)

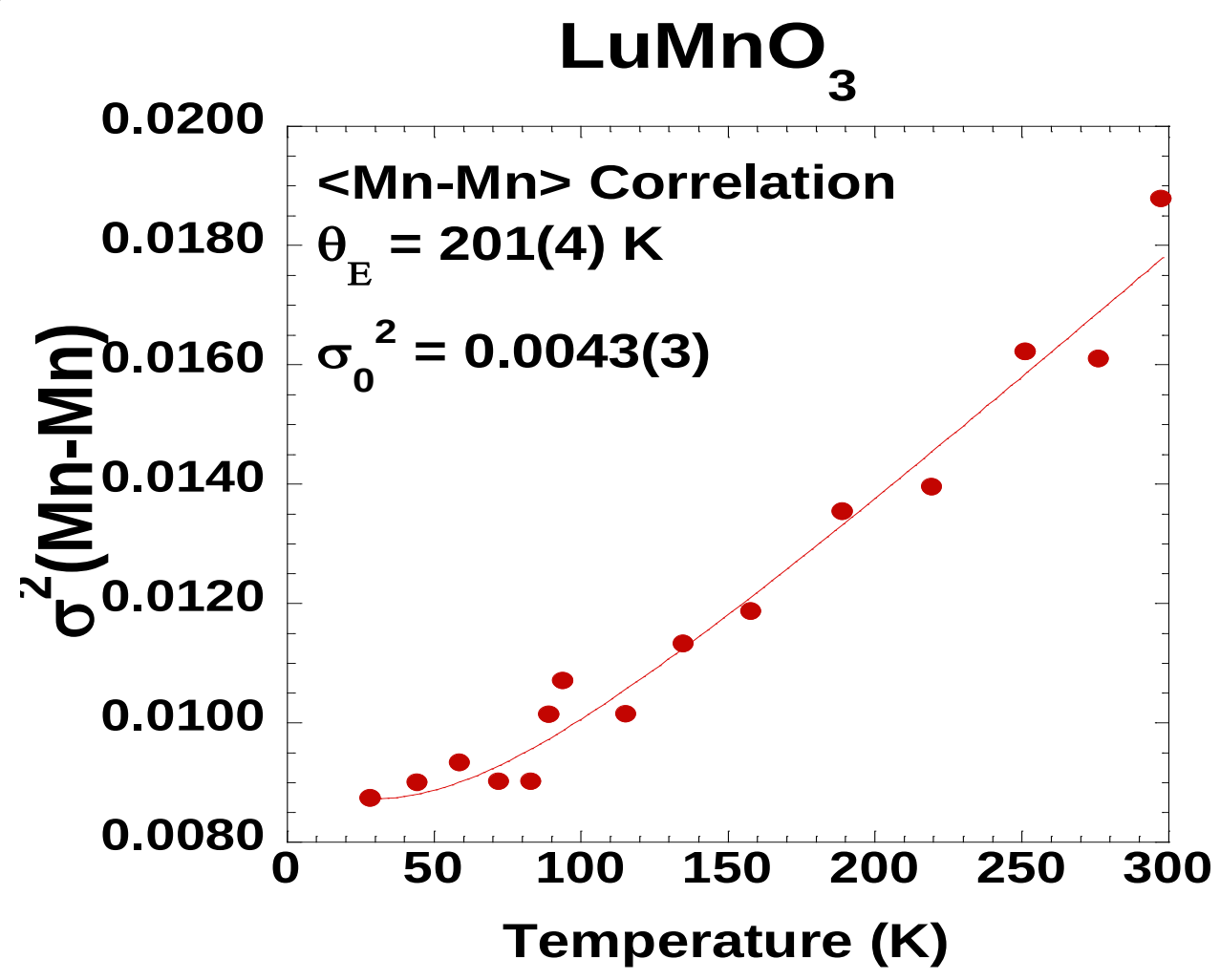

# References

[1] (a) W. Prellier, M. P. Singh, and P. Murugavel,, J. Phys.: Condens. Matter **30**, R803 (2005).

(b) J. S. Zhou, J. B. Goodenough, J. M. Gallardo-Amores, E. Moran, M. A. Alario-Franco, and R. Caudillo, Phys. Rev. B **74**, 014422 (2006).

(c) C. Dubourdieu, G. Huot, I. Gelard, H. Roussel, O. I. Lebedev, and G. Van Tendeloo, Philos. Mag. Lett. **87**, 203 (2007).

[2] (a) C. dela Cruz, F. Yen, B. Lorenz, Y. Q. Wang, Y. Y. Sun, M. M. Gospodinov and C. W. Chu, Phys. Rev. B **71**, 60407R, (2005).

(b) T. Katsufuji, S. Mori, M. Masaki, M. Moritomo, N. Yamamoto and H. Takagi, Phys. Rev B **64**, 104419 (2001).

[3] (a) H. Fukumura, S. Matsui, H. Harima, K. Kisoda, T. Takahashi, T. Yoshimura and N. Fujimura, J. Phys. Cond. Matter **19,** 365239 (2007).

(b) A. P. Litvinchuk, M. N. Iliev, V. N. Popov, and M. N. Gospodinov, J. Phys. Cond. Matter **16**, 809 (2003).

(c) A. B. Souchkov, J. R. Simpson, M.Quijada, H. Ishibashi, H. Hur, J. S. Ahn, S.-W. Cheong, A. J. Millis, and H. Drew, Phys. Rev. Lett. **91**, 27203 (2003).

[4] T. A. Tyson, T. Wu, K. H. Ahn, S.-B. Kim and S.-W Cheong, Phys. Rev. B: Condens. Matter Mater.Phys. **81**, 054101 (2010).

[5] P. Tong, D. Louca, N. Lee and C. W. Cheong, Phys. Rev. B **86**, 094419 (2012).

[6] R. Das, A. Jaiswal, S. Adyanthaya and P. Poddar, J. Phys. Chem. C **114**, 12104 (2010).

[7] T. Yu, P. Gao, T. Wu, T. A. Tyson, and R. Lalancette, Appl. Phys. Lett. **102**, 172901 (2013).

[8] O. V. Dolomanov, L. J. Bourhis, R. J. Gildea, J. A. C. Howard, and H. Puschmann, J. Appl. Cryst. **42**, 339 (2009).

[9] (a) R. B. Neder and Th. Proffen, Diffuse Scattering and Defect Structure Simulations, (Oxford University, Oxford, 2008).

(b) T. Egami and S. L. J. Billinge, Underneath the Bragg Peaks: Structural Analysis of Complex Materials, (Pergamon, Amsterdam, 2003).

(c) Th. Proffen, S. J. L. Billinge, T. Egami and D. Louca, Z. Kristallogr **218**, 132 (2003).

(d) V. Petkov, in Characterization of Materials, (John Wiley and Sons, Hoboken, 2012).

[10] (a) B. Ravel and M. Newville, J. Synchrotron Rad. **12**, 537 (2005).

(b) X-Ray Absorption: Principles, Applications, Techniques of EXAFS, SEXAFS and XANES, edited by D. C. Konningsberger and R. Prins (Wiley, New York, 1988).

[11] J. Mustre de Leon, S. D. Conradson, I. Batistić, A. R. Bishop, I. D. Raistrick, M. C. Aronson, and F. H. Garzon, Phys. Rev. B **45**, 2447 (1992).

[12] (a) A. Muňoz, J.A. Alonso, M.J. Martínez-Lope, M.T. Casáis J.L. Martínez and M.T. Fernández-Díaz, Phys. Rev. B **62**, 9498 (2000).

(b) B. B. Van Aken, A. Meetsma and T. T. M. Palstra, Acta Cryst. E **57**, 101 (2001).

[13] D. G. Tomuta, S. Ramakrishnan, G. J. Nieuwenhuys and J. A. Mydosh, J. Phys. Condens. Mater 13, 4543 (3001).

[14] A. P. Ramirez Ann. Rev. Mat. Sci. **24**, 453 (1994).